\documentclass[lettersize,journal]{IEEEtran}
\usepackage{amsmath,amsfonts}
\usepackage{array}
\usepackage[caption=false,font=normalsize,labelfont=sf,textfont=sf]{subfig}
\usepackage{textcomp}
\usepackage{stfloats}
\usepackage{url}
\usepackage{verbatim}
\usepackage{graphicx}
\usepackage{cite}
\usepackage{caption}
\usepackage{bm,booktabs}
\usepackage{hyperref}
\hypersetup{colorlinks=true,citecolor=blue,linkcolor=blue,filecolor=blue,urlcolor=blue,breaklinks=true}
\DeclareMathOperator{\sgn}{sign}
\DeclareMathOperator{\diag}{diag}
\DeclareMathOperator{\var}{Var}
\usepackage[ruled, vlined, linesnumbered]{algorithm2e}
\begin{document}

\title{Fast Numerical Solver of Ising Optimization Problems via Pruning and Domain Selection}


\author{Langyu Li, Daoyi Dong~\IEEEmembership{Fellow,~IEEE} and Yu Pan~\IEEEmembership{Senior Member,~IEEE}

\thanks{L. Li and Y. Pan are with the College of Control Science and Engineering, Zhejiang University, Hangzhou 310027, China.\\
\indent D. Dong is with the School of Engineering, Australian National University, Canberra ACT 2601, Australia.\\ 
Email address: \{langyuli,ypan\}@zju.edu.cn}
}



\maketitle

\begin{abstract}
Quantum annealers, coherent Ising machines and digital Ising machines for solving quantum-inspired optimization problems have been developing rapidly due to their near-term applications. The numerical solvers of the digital Ising machines are based on traditional computing devices. In this work, we propose a fast and efficient solver for the Ising optimization problems. The algorithm consists of a pruning method that exploits the graph information of the Ising model to reduce the computational complexity, and a domain selection method which introduces significant acceleration by relaxing the discrete feasible domain into a continuous one to incorporate the efficient gradient descent method. The experiment results show that our solver can be an order of magnitude faster than the classical solver, and at least two times faster than the quantum-inspired annealers including the simulated quantum annealing on the benchmark problems. With more relaxed requirements on hardware and lower cost than quantum annealing, the proposed solver has the potential for near-term application in solving challenging optimization problems as well as serving as a benchmark for evaluating the advantage of quantum devices.
\end{abstract}


\section{Introduction}
\IEEEPARstart{A}{s} a fundamental concept in the realm of quantum computing and optimization, the Ising problem has garnered significant attention in recent years. The Ising problem, originally conceived in statistical mechanics~\cite{ising1925contribution}, has found applications in quantum computing~\cite{preskill2018quantum} which is mainly due to its profound potential in solving large-scale combinatorial optimization problems and enhance the performance of machine learning~\cite{lucas2014ising,mcmahon2018solve,weinberg2023supply,bao2021multi,pan2023ising,hussain2020optimal,inoue2021traffic,venturelli2015quantum,parizy2022cardinality,venturelli2019reverse,benedetti2016estimation,glover2006handbook,bohm2022noise}.

\begin{figure*}[!t]
    \centering
    \includegraphics[width=\textwidth]{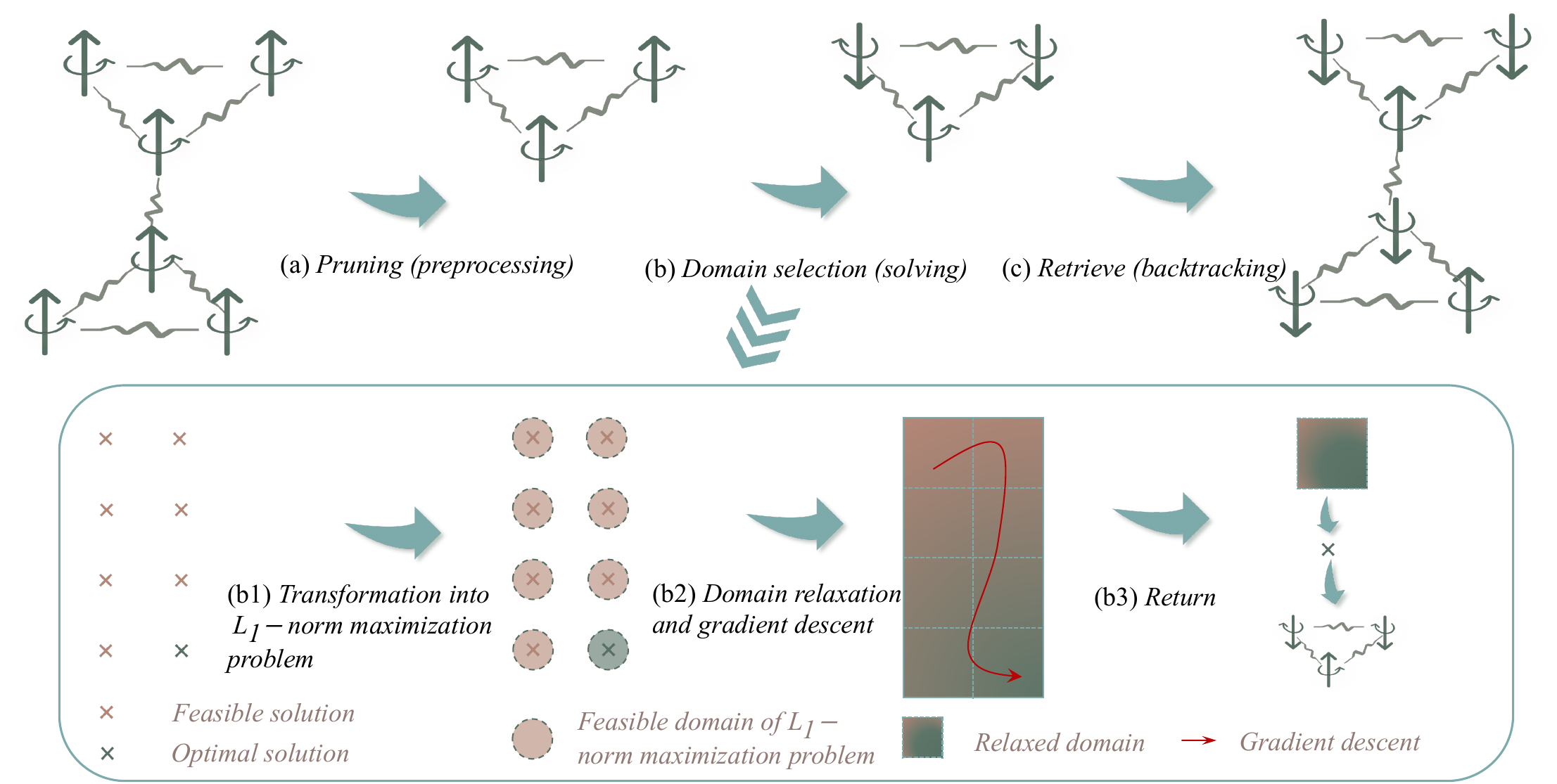}
    \caption{Illustration of our solver. (a) Our solver first applies pruning to simplify the optimization problem by utilizing the graph information of the Ising model. (b) Then the domain selection method is applied to find the approximate optimal solutions. As illustrated in (b1)-(b3), the original problem is firstly transformed into an $L_1$-norm maximization problem with discrete feasible domains. Then we relax the feasible domain to a continuous one and use the gradient descent optimizer to find approximate optimal domain. (c) Finally, the solution of the primal problem is retrieved by combining the results of domain selection and pruning.}
    \label{fig:protocol}
\end{figure*}

Both quantum and classical solvers can be used to solve the Ising optimization problems. Quantum solvers mainly include quantum annealers (QAs), and coherent Ising machines (CIMs) based on superconducting and optical devices, respectively. Grounded on quantum physics and quantum adiabatic theorem, quantum solvers leverage quantum properties such as superposition and entanglement to explore the energy landscapes of complex systems~\cite{farhi2001quantum}. Pioneered by D-Wave, QAs have garnered significant attention due to their potential to outperform classical computing in solving optimization tasks~\cite{johnson2011quantum,albash2018adiabatic}. By employing degenerate optical parametric oscillators, CIMs exploit optical interference phenomena to encode and solve complex Ising problems~\cite{wang2013coherent,inagaki2016coherent,dunjko2016quantum}. Although QAs offer unique capabilities with the inherent quantum parallelism, their severe operating constraints such as low temperature and limited size make it difficult to deploy on small intelligent devices~\cite{ohzeki2020breaking,okada2019improving}. CIMs face similar limitations~\cite{bohm2018understanding}. In contrast, quantum-inspired classical solvers are much cheaper and smaller. Despite the lack of quantum advantages inherent in quantum devices, current classical solvers can achieve speed and solution quality close to that of quantum solvers. Therefore, digital solvers based on traditional devices are more likely to be implemented for realistic applications in the near term. Besides, classical solvers are necessary benchmarks for understanding and measuring the advantage of quantum solvers.

Classical solvers mainly refer to digital Ising machines (DIMs). Although rooted in classical computing, DIMs have shown potential in solving complex Ising optimization problem~\cite{chowdhury2023full,yamaoka201520k,2020A,yamamoto2020statica}. DIMs employ digital hardware to implement the optimization algorithms. The architectures of digital Ising machines can be categorized into non-von Neumann architecture and von Neumann one. The former often requires an application-specific integrated circuit (ASIC) that uses a large number of registers to simulate the spins in the Ising model, and the update of the register states is controlled by the simulated annealing algorithm~\cite{van1987simulated,yamaoka201524,aadit2022massively}. The latter uses the general-purpose computing devices such as CPU, GPU, and field-programmable gate arrays (FPGAs) without physically simulating the spins of the Ising model \cite{wang2024parallel}. In particular, recent works based on non-von Neumann architectures~\cite{aadit2022massively,yamaoka201520k,yamamoto2020statica} were not able to show a significant advantage over the von Neumann architecture~\cite{haribara2016coherent}. 

Among the methods based on von Neumann architecture, the operator-based methods~\cite{rendl2007branch,veszeli2022mean} tend to have high computational complexity and are not suitable for large-scale problems. Meanwhile, the accuracy and speed of the other numerical solvers~\cite{bowles2022quadratic,li2023simulated} still have room for improvement. In this work, we propose a fast and scalable solver shown in Fig.~\ref{fig:protocol}, which demonstrates an advantage in speed and accuracy. This solver first applies pruning to simplify and reduce the dimensions of the Ising optimization problem by pre-determining the directions of certain spins. After this preprocessing, the domain selection method will solve the simplified problem to find the approximate optimal solutions. In contrast to previous works, the main innovation is the fusion of the discrete programming method with the gradient descent. To be more specific, we transform the Ising problem into an $L_1$-norm maximization problem, which is then decomposed into several sub-linear programming problems. The feasible sets between these linear programming problems are not continuous, and the optimal solution of each sub-problem is easy to find. Therefore, we find a connection between the energy of the feasible solution and the feasible domain, and design loss functions weighted by the energy of the feasible solution. By making the feasible domain continuous, gradient descent can be used to quickly find a feasible domain containing a near-optimal solution, which is the optimal solution in this domain. In the last, retrieve the solutions of the primal Ising problem from the solution of domain selection with the information recorded in pruning.

A series of experiments have been conducted to evaluate the performance of our solver. In particular, our solver has been extensively compared with Gurobi, D-Wave classical  steepest descent solver, simulated annealing and simulated quantum annealing. Our solver is at most $10$ times faster than Gurobi, which is the state-of-the-art solver based on operation programming. Our solver is also significantly faster than other classical numerical methods and quantum simulated annealing. The comparisons are made on randomly generated samples, as well as some of the most famous NP-hard optimization problems such as maximum cut. The computational complexity has also been verified with the experiment data, which demonstrates the outstanding scalability of our solver as the problem size scales up.

This paper is organized as follows. Sec.~\ref{sec:ising problem} introduces the mathematical model of Ising optimization problem. Sec.~\ref{sec:pruning and domain selection} describes our solver, with Sec.~\ref{sec:pruning} and Sec.~\ref{sec:domain selection} introducing the pruning and domain selection algorithms, respectively. Sec.~\ref{sec:numerical result} summarizes the experiment results and comparison. Sec.~\ref{sec:conclusion} concludes this work.

\section{Ising problem}\label{sec:ising problem}

\begin{figure}[!t]
    \centering
    \includegraphics[width=\linewidth]{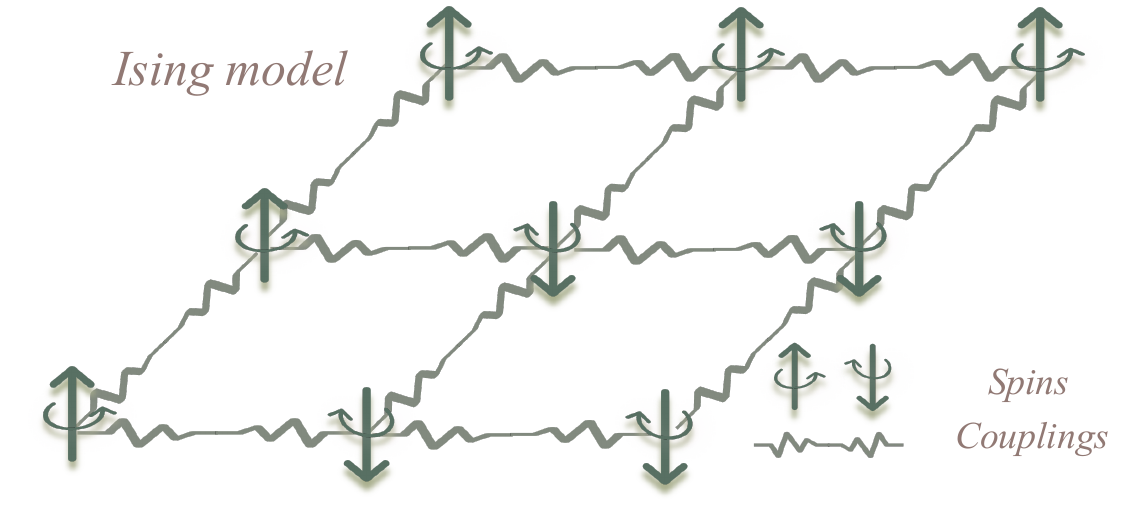}
    \caption{A lattice Ising model without external field.}
    \label{fig:ising model}
\end{figure}

The Ising model was first formulated by Ernst Ising in 1925~\cite{ising1925contribution}, which was initially conceived as a simple model for describing magnetic interactions in crystalline systems explaining the behavior of spins interacting with their neighbors. The Hamiltonian of the Ising model is defined by 
\begin{align}\label{eq:ising hamiltonian}
    H = \sum_{i>j} \widehat{J}_{ij} \sigma_i^z \sigma_j^z + \sum_i \widehat{h}_i \sigma_i^z,
\end{align}
where $\sigma_i^z$ is the Pauli-$Z$ operator acting on the $i$-th qubit. The coupling amplitude $\widehat{J}_{ij}\in \mathbb{R}$ is a scalar to describe the strength of the coupling between the qubits $i$ and $j$, and the coefficient $\widehat{h}_i \in \mathbb{R} $ describes the external field strength applied on the $i$-th qubit. The qubits are also called spins since each qubit has two states $|0\rangle $ and $|1\rangle$ corresponding to spin up and spin down, respectively. Fig.~\ref{fig:ising model}
shows a lattice Ising model.
The state of the Ising model can be described by a state vector $|\psi\rangle \in \{ |0\rangle, |1\rangle\}^{\otimes d}$. For a $d$-qubit Ising model with the state vector $|\psi\rangle$, its energy, or called landscape, is given as
\begin{align}\label{eq:observation}
    \langle H \rangle = \langle \psi | H|\psi \rangle, \quad |\psi \rangle \in \{ |0\rangle, |1\rangle\}^{\otimes d}.
\end{align}
Eq.~\eqref{eq:observation} has an equivalent scalar form as
\begin{align}
\langle H \rangle &\cong E(\bm{s}),\\
    E(\bm{s}) &=\bm{s}^\top \bm{J}\bm{s} + \bm{h}^\top \bm{s},\quad \bm{s} \in \{ -1, 1\}^d,
\end{align}
where $\bm{s}$ is a $d$-dimensional vector called spin configuration with respect to the state vector $|\psi\rangle$. The $i$-th entry of $\bm{s}$ is $+1$ if the $i$-th spin is at the state $|0\rangle$, and $-1$ if the spin state is $|1\rangle$. Coupling matrix $\bm{J}\in \mathbb{R}^{d\times d}$ is defined by $J_{ij} = J_{ji}=\frac{1}{2}\widehat{J}_{ij}$, which is real symmetric with $J_{ii}=0$. Field vector $\bm{h}\in \mathbb{R}^d$ is defined by $h_i = \widehat{h}_i$. The flip operator on the $i$-th spin is defined as 
\begin{align}
    X_i = \diag (1,\cdots,1,\underset{i\text{-th}}{-1},1,\cdots,1).
\end{align}
We use the notation $X_{i,j} = X_i X_j$ since $X_i$ and $X_j$ commute.

The Ising optimization problem refers to the task of finding the optimal configuration of spins within a system to minimize or maximize the given objective function. In particular, most of the Ising optimization problem can be formulated as finding the ground-state energy of the Ising model. Mathematically, we can define this primal problem as the following quadratic unconstrained binary optimization (QUBO)
\begin{align}\label{eq:primal}
    \min_{\bm{s}} \quad &E(\bm{s}),\\
    \text{s.t.}\quad  &\bm{s}\in \{ -1,1\}^d.
\end{align}
$\bm{s}^*$ denotes the ground state vector or optimal state vector, which is written as  
\begin{align}
    \bm{s}^* = \arg \min_{\bm{s}} \quad &E(\bm{s}),\quad\bm{s}\in \{ -1,1\}^d.
\end{align}

\section{Pruning and Domain Selection}\label{sec:pruning and domain selection}

In this section, we will introduce the numerical solver which consists of two stages, namely pruning and domain selection. Pruning can be considered as a preprocessing step of the primal Ising problem based on the graph information, and domain selection is an efficient optimization method to produce a near-optimal solution. 

\subsection{Pruning}\label{sec:pruning}
Pruning is the process of pre-determining the optimal directions of certain spins. For an arbitrary state $\bm{s}$, its energy difference between an arbitrarily flipped spin configuration $\bm{s}^{\prime}$ is calculated as
\begin{align}
     &E(\bm{s}^{\prime}) -E(\bm{s})= E(X_\mathcal{F} \bm{s}) -E(\bm{s})\\
    =& \bm{s}^\top (X_\mathcal{F} \bm{J} X_\mathcal{F} -\bm{J}) \bm{s}+(\bm{h}^\top X_\mathcal{F} - \bm{h}^\top )\bm{s}\\
    =&-4\sum_{i\in \mathcal{F}} s_i \sum_{j=1}^d J_{ij}s_j-2\sum_{i\in \mathcal{F}}h_i s_i,
\end{align}
where $X_\mathcal{F} = \prod_{i\in \mathcal{F}} X_i$. For the $i$-th spin, if there holds $2\sum_j |J_{ji}| < |h_i| $, we must have
\begin{align}\label{eq:determined spin}
    s^*_i = - \sgn (h_i),
\end{align}
since the energy of the configuration will definitely increase if $s^*_i$ is flipped
\begin{align}
 &E(X_i \bm{s}) -E(\bm{s}) \\
 =& -4 s_i \sum_{j=1}^d J_{ij}s_j - 2 h_i s_i  \\
    =& -4 s_i \sum_{j=1}^d J_{ij}s_j + 2 \sgn (h_i)h_i  \\
    \geq & 2 |h_i| -4 \big |s_i  \sum_{j=1}^d J_{ij}s_j \big|> 0. 
\end{align}
This class of spins are named as determined spins. Similarly, if the $i$-th spin is coupled to the $j$-th spin only and there holds $2|J_{ij}| \geq |h_i|$, we must have
\begin{align}\label{eq:semi determined spin}
    s^*_i = -s^*_j\sgn(J_{ij}).
\end{align}
This class of spins are named as semi-determined spins since their directions are solely dependent on the neighboring spins. 

Therefore, given an Ising problem, the spins can be divided into determined spins $\mathcal{D}$, semi-determined spins $\mathcal{S}$, and other spins $\mathcal{G}$. Without loss of generality, we write the simplified optimization problem after a single pruning as
\begin{align}\label{eq:simplified optimization}
   \min_{\bm{s}_{\mathcal{G}}} \quad & \bm{s}^\top_{\mathcal{G}} \bm{J}_{\mathcal{G}}\bm{s}_{\mathcal{G}} + \bm{h}^\top_{\mathcal{G}} \bm{s}_{\mathcal{G}} +C_{\mathcal{D},\mathcal{S}} ,\\
   \text{s.t.}\quad  &\bm{s}_{\mathcal{G}}\in \{ -1,1\}^{|\mathcal{G}|},   
\end{align}
where
\begin{align}
    (\bm{J}_{\mathcal{G}})_{ij} =& J_{ij} , \quad\text{for $i,j \in \mathcal{G}$} ,\\
    (\bm{h}_{\mathcal{G}})_{i} 
    =& h_i - 2\sum_{j\in \mathcal{D}} J_{ij} \sgn(h_j)   - \sum_{k\in \mathcal{S}}  \sgn(J_{ik}  )  h_k \\
    &+ 2\sum_{k\in \mathcal{S}}\sum_{j\in \mathcal{D}} J_{kj}  \sgn(J_{ik} h_j),\quad \text{for $i \in \mathcal{G}$},\\ 
    C_{\mathcal{D},\mathcal{S}} 
    =& -\sum_{i\in \mathcal{D} } |h_i|   - 2\sum_{j\in \mathcal{G}}\sum_{k\in \mathcal{S}}|J_{jk}| + \sum_{i,j\in \mathcal{D}} J_{ij} \sgn(h_i h_j). \label{eq:simplified optimization end}
\end{align}
The pruning algorithm can then be defined recursively according to Algorithm~\ref{alg:pruning}. 


\begin{algorithm}
\caption{Pruning}
\label{alg:pruning}
\SetAlgoLined
\SetKwInOut{Input}{input}\SetKwInOut{Output}{output}
\Input{$d$-spin Ising problem with coupling matrix $\bm{J}$ and field vector $\bm{h}$}
\Output{A $d^\prime$-spin Ising problem with matrix $\bm{J}^\prime$ and vector $\bm{h}^\prime$,$d^\prime \leq d$}

\While{$\mathcal{S}\cup \mathcal{D} \neq \emptyset $}{

Select the determined spins into $\mathcal{D}$;

Select the semi-determined spins into $\mathcal{S}$;

Divide the current spins into $\mathcal{G},\mathcal{S},\mathcal{D}$;

Redefine the current Ising problem by Eq.~\eqref{eq:simplified optimization}~$\sim$~\eqref{eq:simplified optimization end}.
}
\Return The simplified Ising problem with $\bm{J}^\prime$ and $\bm{h}^\prime$.
\end{algorithm}


\subsection{Domain Selection}\label{sec:domain selection}
The domain selection method first converts the primal problem into an $L_1$-norm maximization problem whose feasible domain is not continuous. Then the feasible domain is relaxed to a continuous one to facilitate gradient-based optimization method.

By introducing another variable we can transform the primal problem into 
\begin{align}\label{eq:two variable problem}
    \min_{\bm{s}_x, \bm{s}_y} &\quad \bm{s}_x^\top \bm{J} \bm{s}_y + \bm{h}^\top \bm{s}_y,\\
    \text{s.t.}&\quad \bm{s}_x  = \bm{s}_y,\\
    &\quad \bm{s}_x  ,\bm{s}_y  \in \{ -1, 1\}^d.
\end{align}
The following metric function $D(\bm{s}_x, \bm{s}_y)$ 
\begin{align}
    D(\bm{s}_x, \bm{s}_y) = d- \bm{s}_x^\top \bm{s}_y
\end{align}
is used to measure the distance between $\bm{s}_x$ and $\bm{s}_y$. Thus the augmented Lagrange formulation of this two-variable problem is given by 
\begin{align}\label{eq:lagrange function}
  \min_{\bm{s}_x, \bm{s}_y} &\quad \bm{s}_x^\top \bm{J} \bm{s}_y + \bm{h}^\top \bm{s}_y + \mu ( d- \bm{s}_x^\top \bm{s}_y ),\\
    \text{s.t.}&\quad \bm{s}_x  ,\bm{s}_y  \in \{ -1, 1\}^d,
\end{align}
where $\mu >0$ is a weight scalar. The optimal solution to this problem can be calculated as
\begin{align}
&\min_{\bm{s}_x, \bm{s}_y} \quad \bm{s}_x^\top \bm{J} \bm{s}_y + \bm{h}^\top \bm{s}_y - \mu  \bm{s}_x^\top \bm{s}_y \\
=& \min_{\bm{s}_x, \bm{s}_y}\quad (\bm{s}_x^\top \bm{J}  + \bm{h}^\top  - \mu  \bm{s}_x^\top )\bm{s}_y \\
 =&\min_{\bm{s}_x} \min_{\bm{s}_y}\quad (\bm{s}_x^\top \bm{J}  + \bm{h}^\top  - \mu  \bm{s}_x^\top )\bm{s}_y \label{eq:1 norm 1}\\
 =&\min_{\bm{s}_x} \quad -\Vert \bm{J} \bm{s}_x  + \bm{h}  - \mu  \bm{s}_x \Vert_1 \label{eq:1 norm 2}\\
=&\max_{\bm{s}} \quad \Vert (\bm{J}-\mu  I) \bm{s}  + \bm{h}  \Vert_1. \label{eq:1 norm 3}
\end{align}
Note that in Eq~\eqref{eq:1 norm 1}$\sim$\eqref{eq:1 norm 2} the optimal $\bm{s}_y^*$ is given by $(\bm{s}_y^*)_i = -\sgn((\bm{s}_x^\top \bm{J}  + \bm{h}^\top  - \mu  \bm{s}_x^\top )_i)$ and $\Vert \cdot \Vert_1$ is the $L_1$ norm. A large enough $\mu$ can guarantee that the solution to the Augmented Lagrangian problem is the same as the solution to the primal problem.

The $L_1$-norm maximization defined in \eqref{eq:1 norm 3} can then be decomposed into $2^d$ linear programming problems. Then the $L_1$-norm optimization problem has an equivalent form of
\begin{align}
 \max_{\bm{w}} &\quad \bm{1}^\top \bm{w},\\ 
  \text{s.t.}& \quad \bm{w} \in \mathcal{C},
\end{align}
where $\bm{1}$ is the all-one vector. The feasible set of solutions is the concatenation of $2^d$ convex sets, which is given by
\begin{align}
    \mathcal{C}=& \bigcup_{i=0}^{2^d-1} \mathcal{C}_i,
\end{align}
and each feasible set $\mathcal{C}_i$ is constrained by
\begin{align}
    \mathcal{C}_i = \{ \bm{w}: \bm{w} + X_{\{i\}}((\bm{J}-\mu I)\bm{s} +\bm{h}) \leq 0, \bm{s}  \in \{ -1, 1\}^d \},
\end{align}
with $X_{\{i\}}= \prod_{j\in \text{idx}(i)} X_j $. $\text{idx}(i)$ returns the index set of $d$-dimension binary vector whose entry value is equal to 1, and the binary vector can be seen as a binary number whose value in decimal is $i$. Consequently, the $L_1$-norm optimization problem becomes a feasible domain search for the $\mathcal{C}_i$ where the optimal solution locates. Finding such an optimal value in a certain feasible domain is a mixed-integer linear programming problem, which has exponential complexity in principle~\cite{bulut2021complexity}. In our case, this problem is relatively easy to solve since the local optimal solution of $\mathcal{C}_i$ can be obtained as
\begin{align}\label{eq:local optimal}
  \widetilde{\bm{s}}^*=- \sgn(X_{\{i\}} (\bm{J}-\mu I)\bm{1}) =\sgn(X_{\{i\}}\bm{1}) .  
\end{align}
Therefore, all we need to do is to find the optimal domain $\mathcal{C}_i^*$ if we encode the local optimal solutions into the feasible domains such that different feasible domains have different costs. The discrete $X_{\{i\}}$ can be relaxed by multiplying a factor of matrix function $\sin(\bm \theta)$ as 
\begin{align}
 \max_{\bm{w}} &\quad \bm{1}^\top \bm{w},\\ 
  \text{s.t.}& \quad  \bm{w} + \sin{(\bm{\theta} )}((\bm{J}-\mu I)\widetilde{\bm{s}}^* +\bm{h}) \leq 0.
\end{align}
where $\bm{\theta} = \diag \{ \theta_1, \theta_2,\cdots, \theta_d\}$. Such a relaxation transforms the original discrete parameter space into a continuous parameter space. By incorporating the constraints into the optimization target, the loss function for domain selection finally becomes 
\begin{align}\label{eq:loss function}
    \mathcal{L} (\bm{\theta}) = \bm{1}^\top \sin{(\bm{\theta})} ((\bm{J}-\mu I)\widetilde{\bm{s}}^* +\bm{h}).
\end{align}
With the loss function defined in Eq.~\eqref{eq:loss function}, the domain selection can be efficiently solved with any gradient descent method by optimizing $\bm{\theta}$. 
\section{Numerical Results}\label{sec:numerical result}

\subsection{Pruning Rate}
To evaluate the performance of the pruning algorithm, the pruning rate $\eta$ is defined as
\begin{align}\label{eq:pruning rate}
    \eta := 1-\frac{d_\mathcal{G}}{d},
\end{align}
where $d_\mathcal{G} = |\mathcal{G}|$ is the number of spins that are not pruned. Let $\mathcal{U}, \mathcal{N}, \mathcal{B}$ denote uniform, Gaussian, and uniformly binary distributions, respectively. For a $d$-qubit $k$-regular Ising optimization problem with $k\leq d-1, \widehat{J}_{ij} \sim \mathcal{U}(-\alpha,\alpha)$ and $\widehat{h}_i \sim \mathcal{U}(-\beta,\beta)$, the pruning rate $\eta_1$ of the first round can be estimated as
\begin{align}
\eta_1&=\Pr (|\widehat{h}_i| > \sum_ j|\widehat{J}_{ij}|)\label{eq:bound expectation 1}\\
     &\approx \Pr\big(b>a:a\sim \mathcal{N}\big(\frac{k\alpha}{2}, \frac{k\alpha^2}{12}\big),b\sim \mathcal{U}(0,\beta)\big)\label{eq:bound expectation 2}\\
     &= \int_0^\beta \frac{1}{\beta} ~\text{d}x  \int_{-\infty}^x \frac{\sqrt{6}}{\sqrt{\pi k}\alpha } \exp(-6\frac{(y-\frac{k\alpha}{2})^2}{k\alpha^2}) ~\text{d}y,\label{eq:bound expectation}
\end{align}
where we have used the Gaussian distribution to approximate the Irwin–Hall distribution~\cite{johnson1995continuous} in Eq.~\eqref{eq:bound expectation 1} according to the central limit theorem.

\subsection{Performance}
Here we introduce the Hamiltonian ratio $\chi$ as a normalized metric to evaluate the Hamiltonian error with respect to the ground state~\cite{li2023simulated}, which is defined as 
\begin{align}\label{eq:hamiltonian ratio}
    \chi &=  \frac{E_{\max}-\widehat{E}}{E_{\max} - E_{\min}} ,
\end{align}
where $\widehat{E}$ is the solution, and $E_{\max}$ and $E_{\min}$ are the maximal and minimal energies of the Ising model, respectively. Due to the difficulty in obtaining the exact $E_{\max}$ and $E_{\min}$, $E_{\max}$ and $E_{\min}$ are estimated by
\begin{align}
   \widehat{E}_{\max} =-\widehat{E}_{\min}=\sqrt{d} \Vert \bm{J} \Vert_F + \Vert \bm{h} \Vert_1,
\end{align}
where $\Vert \cdot \Vert_F$ is the Frobenius norm. The Hamiltonian ratio is thus approximated by
\begin{align}
    \widehat{\chi} &=  \frac{\widehat{E}_{\max}-\widehat{E}}{\widehat{E}_{\max} -\widehat{E}_{\min}} =\frac{1}{2}-\frac{\widehat{E}}{2(\sqrt{d} \Vert \bm{J} \Vert_F + \Vert \bm{h}\Vert_1)}.
\end{align}
Such an approximation effectively normalizes the Hamiltonian ratio, and $\widehat{\chi}$ is linearly and positively correlated with $\chi$ as 
\begin{align}
    \chi = \frac{E_{\max}-(1-2  \widehat{\chi} )(\sqrt{d} \Vert \bm{J} \Vert_F + \Vert \bm{h} \Vert_1)}{E_{\max} - E_{\min}}. 
\end{align}

\begin{table*}[!t]
\centering
\caption{The performance of pruning and domain selection. PDS and DS refer to domain selection with and without pruning, respectively. Iterations are the number of steps taken to reach the convergence. $T$ is the time cost in seconds. $\Delta \widehat{\chi} =\widehat{\chi}_{\text{PDS}} - \widehat{\chi}_{\text{DS}} $.}
    \label{tab:performance}

    \resizebox{\textwidth}{!}{\begin{tabular}{c|c c c |c c c |c c}
    \toprule
         &DS Iterations  & $T_{\text{DS}} $ & $\widehat{\chi}_{\text{DS}}$&PDS Iterations&$T_{\text{PDS}} $ &$\widehat{\chi}_{\text{PDS}}$ & $\eta$&$\Delta \widehat{\chi}$ \\
         \hline
         Uniform&$89.5 \pm 7.4 $ & $0.279\pm 0.035$& $0.9248 \pm 0.0037$  & $74.7\pm 12.5$  & $0.056\pm0.013$& $0.9261\pm 0.0036$  &$0.949\pm 0.018$    & $0.0014 \pm 0.0004$\\
         Gaussian&$94.2\pm 3.4 $ & $0.297\pm 0.020$& $0.9142 \pm 0.0052$  & $90.4\pm 5.3$  & $0.246\pm 0.018$& $0.9145\pm 0.0051$  &$0.290\pm 0.030$    & $0.0003 \pm 0.0007$\\
         Binary&$92.7\pm5.4 $ & $0.289\pm 0.028$& $0.8815\pm 0.0049$  & $93.0\pm 5.2$  & $0.296\pm 0.025$& $0.8815\pm 0.0049$  &$0.0$    & $0.0000\pm 0.0005$\\
    \bottomrule
    \end{tabular}}
\end{table*}

\begin{table*}[!t]
    \centering
    \caption{Details of the experiments in Fig.~\ref{fig:compare}.}
    \label{tab:comparisons}
    \resizebox{0.8\textwidth}{!}{
    \begin{tabular}{c|cccc cccc}
    \toprule
         & $\mathbb{E}(\widehat{\chi})$ &$\var (\widehat{\chi})$ & $\mathbb{E}(T)$ &$\var (T)$ &$\mathbb{E}(\widehat{T}_{\text{TTS}})$ &$\var (\widehat{T}_{\text{TTS}})$ &$\mathbb{E}(v)$ &$\var (v)$  \\
         \hline
         Ours&1.169&0.002&\textbf{0.196}&0.001 &\textbf{280.6}&2183.5&\textbf{6.041}&0.393\\ 
         Gurobi&\textbf{1.209}&0.003&3.373&0.103&4641.2 &377051.4&0.361&0.001\\ 
         DWave-SD&1.149&0.002&0.319&0.002 &465.9&5880.7&3.625&0.057\\ 
         OpenJij-SA&1.115&0.002&2.127&0.039 &3208.0&102109.2&0.526&0.001\\ 
         OpenJij-SQA&0.972&0.001&2.181&0.001 &3793.2&40407.6&0.447&0.001\\ 
    \bottomrule
    \end{tabular}}
\end{table*}

\begin{table*}[!htp]
    \centering
    \caption{Details of the experiments in Fig.~\ref{fig:typical}.}
    \label{tab:typical}
    \resizebox{\textwidth}{!}{
    \begin{tabular}{c|cccc |cc|cc|cc|cc|cc}
    \toprule
          &&&& &\multicolumn{2}{c|}{Ours}&\multicolumn{2}{c|}{Gurobi}&\multicolumn{2}{c|}{DWave-SD} &\multicolumn{2}{c|}{OpenJij-SA}&\multicolumn{2}{c}{OpenJij-SQA} \\
         Problem &$J_{ij}$ &$h_{i}$ & density &$k$-regular &$T$ &$\widehat{E}$&$T$ &$\widehat{E}$&$T$ &$\widehat{E}$&$T$ &$\widehat{E}$&$T$ &$\widehat{E}$ \\
         \hline
         MaxCut-3&$\{0,1\}$&0&0.003&3&0.23&-2312&\textbf{0.16}&\textbf{-2408}&0.33&-2188&2.12&-2176&2.29&-1987\\ 
         MaxCut-D&$\{0,1\}$&0&0.5&/&\textbf{0.24}&-22552&3.92&\textbf{-23904}&0.34&-20944&2.14&-20616&2.28&-16152\\ 
         SK-Ising&$\{-1,1\}$&0&1&complete&\textbf{0.21}&-44276&3.65&\textbf{-46280}&0.34&-43708&2.09&-39904&2.24&-31388\\ 
         SK-QUBO&$\{0,1\}$&0&1&complete&\textbf{0.18}&-22098&4.10&\textbf{-23982}&0.33&-22094&2.15&-20266&2.09&-15850\\ 
         NAE-3SAT&$\mathbb{Z}$&0& 0.0125&/&\textbf{0.22}&-4012&0.95&\textbf{-4052}&0.33&-3388&2.11&-3668&2.25&-3268\\ 
    \bottomrule
    \end{tabular}}
\end{table*}

We use the $1000$-spin $6$-regular Ising models for the performance demonstration. The parameters of the Ising models are randomly generated according to the uniform distribution $\widehat{J}_{ij} \sim \mathcal{U}(-1,1),\widehat{h}_i \sim \mathcal{U}(-6,6)$, Gaussian distribution $\widehat{J}_{ij} \sim \mathcal{N}(0,\frac{1}{3}),\widehat{h}_i \sim \mathcal{N}(0,12)$, and binary distribution $\widehat{J}_{ij} \sim \mathcal{B}(\{-\frac{1}{\sqrt{3}},\frac{1}{\sqrt{3}}\}),\widehat{h}_i \sim \mathcal{B}(\{-2\sqrt{3},2\sqrt{3}\})$. These three distributions have the same mean and variance. The experiment results are shown in Tab.~\ref{tab:performance}, where we have repeated $1000$ trials for each distribution. ADAM optimizer~\cite{kingma2014adam} is used with a learning rate of $6.0$ for the gradient descent. The condition for convergence is set as
\begin{align}
\frac{\Delta\mathcal{L}(\bm{\theta})}{\mathcal{L}(\bm{\theta}_\text{now})-\mathcal{L}(\bm{\theta}_\text{init})} \leq 0.005,
\end{align}
where $\bm{\theta}_\text{now},\bm{\theta}_\text{init}$ are the current and initial $\bm{\theta}$. $\Delta\mathcal{L}(\bm{\theta})$ is the difference between the maximum and the minimum of the losses in the last 5 iterations. $\bm{\theta}_\text{init}$ is randomly initialized. For the uniform model, the pruning rate is about $95\%$, which leads to a significant reduction in computing time. The pruning also has a positive effect on the computing time for Gaussian model. but the acceleration is smaller due to the smaller pruning rate. 

\subsection{Comparisons}
Here we use the time-to-solution (TTS) as a benchmark~\cite{king2015benchmarking}
\begin{align}
    T_{\text{TTS}}   = T \frac{\log (1- 0.99)}{\log (1- P)},
\end{align}
where $P$ is the success probability of finding the optimal value in one run and $T$ is the time cost of this run. Generally speaking, $P$ is difficult to calculate since the optimal solution is unknown. Hence we use the estimation $P=\frac{1}{d} e^{\chi}$ since the time complexity tends to be exponential~\cite{hamerly2019experimental}. The estimated time-to-solution is then given by 
\begin{align}
    \widehat{T}_{\text{TTS}} = T \frac{\log (1- 0.99)}{\log (1- \frac{1}{d} e^{\widehat{\chi}})} .
\end{align}
Apart from TTS,  the solution speed can also be simply defined as 
\begin{align}
    v = \frac{\widehat{\chi}}{T},
\end{align}
which is taken as another metric of efficiency.

\begin{figure}[!htbp]
    \centering
    \subfloat[Comparisons on Key Indicators \label{fig:compare 1}]{\includegraphics[width=\linewidth]{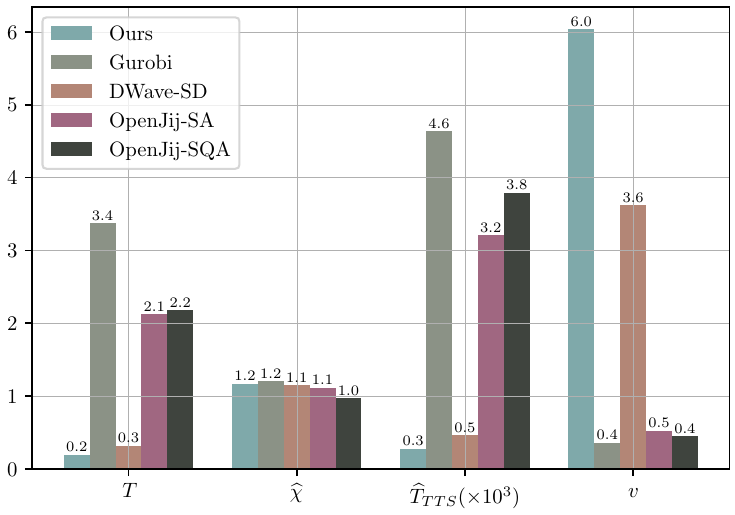}}\\
    \subfloat[Statistical Visualization of Sample Points \label{fig:compare 2}]{\includegraphics[width=\linewidth]{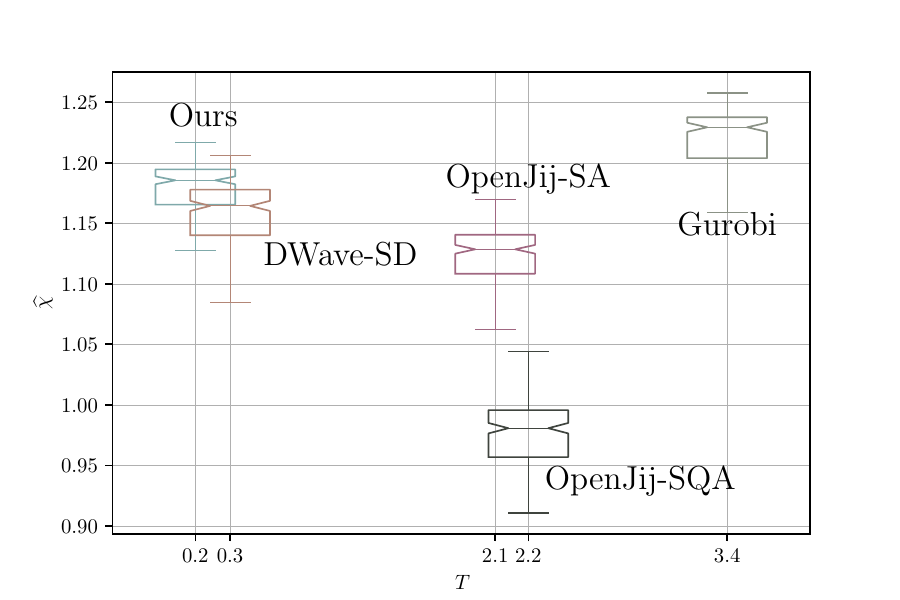}}
    \caption{Comparisons of different solvers on the randomly generated problems. We sample $5000$ complete Ising problems of $1000$ spins. Except for Gurobi, the number of sweeps is set to be $5$. For Gurobi, we set the gap of mixed-integer programming as $\texttt{MIPGap}=0.8$. Running time is measured in seconds.}
    \label{fig:compare}
\end{figure}

We compare our solver with the Python release of Gurobi \texttt{v}$10.0.1$, D-Wave classical solver of steepest descent (DWave-SD), OpenJij with simulated quantum annealing (OpenJij-SQA) and simulated annealing  (OpenJij-SA). All the experiments are conducted at the Apple M2 Pro core. To simulate the models with broader parameters, we generate the optimization problems using complete graphs and the random parameters from the distributions $J_{ij},h_i \sim \mathcal{U}(-1000,1000)$ and $J_{ij},h_i \sim \mathcal{N}(0,\alpha^2),\alpha \sim \mathcal{U}(0,1000)$.  Comparison results are shown in Fig.~\ref{fig:compare}, and more details can be found in Tab.~\ref{tab:comparisons}.

From Fig.~\ref{fig:compare 1} and Tab.~\ref{tab:comparisons} we can see that our solver has the shortest solution time in terms of TTS and $v$. In particular, TTS is a comprehensive metric that characterizes the solution accuracy, the solution time, and the size of the problem, which means that our solver has the potential to find the global optimal solution in the least time. The solution accuracy of our solver is the second highest after Gurobi. Since the operation programming method implemented in Gurobi often requires a large number of operations, its computational complexity is high and the variance of its running time is significantly higher than those of the other methods as can be seen in Tab.~\ref{tab:comparisons}. Compared with the second fastest DWave-SD, our solver still has a clear advantage. In the last, simulated annealing and quantum annealing do not show any advantage over our solver in terms of performance, with worse solution accuracy and a running time which is $9$ times more than that of ours.

\begin{figure}[!htp]
    \centering
    \includegraphics[width=\linewidth]{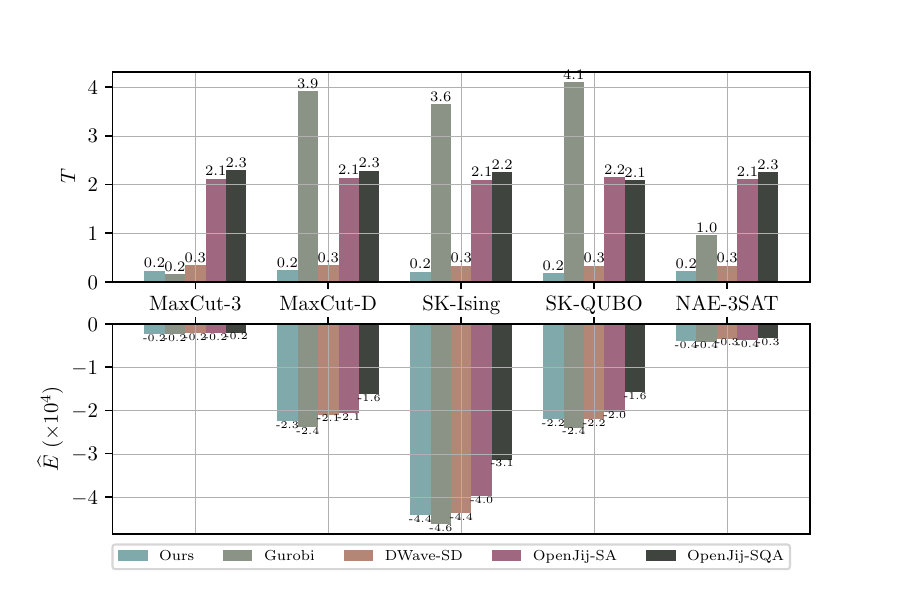}
    \caption{Comparisons of different solvers on the typical fundamental optimization problems.}
    \label{fig:typical}
\end{figure}

To further illustrate the performance of our solver, we conduct experiments on the typical benchmark optimization problems with the same settings, including MaxCut-3 and MaxCut-D~\cite{oshiyama2022benchmark}. The Sherrington–Kirkpatrick (SK) model is an Ising spin glass model with infinite spatial dimensions. We consider the SK spin-glass model (SK-Ising) on a fully connected graph where the couplings $J_{ij} = \pm 1$ are randomly chosen with equal probability, and SK-QUBO model on a fully connected graph with the couplings $J_{ij} = 0,1$~\cite{hamerly2019experimental}. The satisfiability problem (SAT) is one of the most fundamental NP-hard problems which serves as a good benchmark for heuristic solvers. Not-All-Equal 3-SAT (NAE-3SAT) is a variant of the Boolean SAT problem~\cite{oshiyama2022benchmark} for which we have used randomly generated instances with the clause-to-variable ratio $2.11$ for benchmarking. The experiment results can be found in Tab.~\ref{tab:typical} and Fig.~\ref{fig:typical}. 

First, we need to mention that pruning algorithms basically do not work on these models with integer parameters. Our solver still holds the lead both in accuracy and speed over the numerical and simulated quantum annealing methods. Gurobi is more accurate than our solver. However, it can be seen that the speed advantage of our solver over Gurobi increases as the problem densities increase. For highly dense problems, the speed of our solver is generally $16$ times faster than that of Gurobi.

\begin{figure}[!htp]
    \centering
    \includegraphics[width=\linewidth]{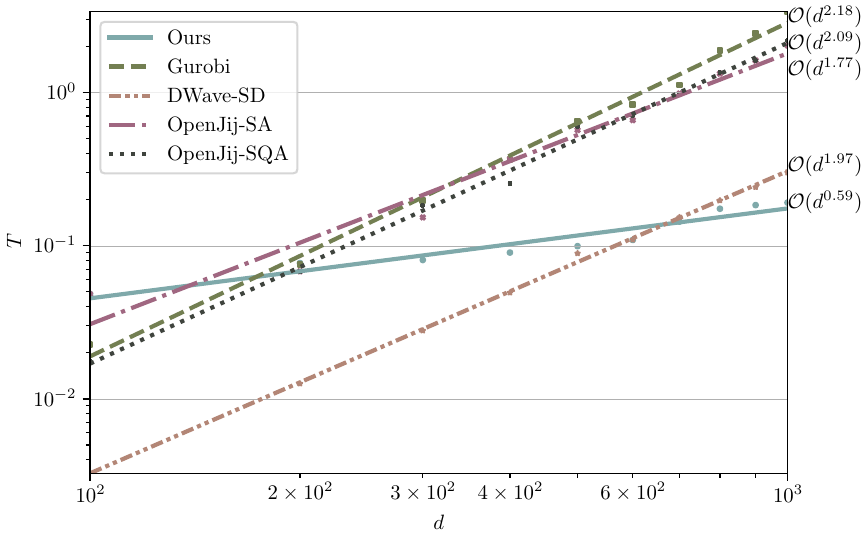}
    \caption{Comparisons of different solvers on scalability. Here we show the variations of $T$ with the number of spins $d$. Apparently, the logarithmic horizontal and vertical axes can be related by polynomial functions. The slope of the line is the highest power of the terms of the polynomial, whose value is labeled on the right of the figure.}
    \label{fig:scale}
\end{figure}

We have also explored the scalability of each solver by varying the size of the Ising problems. The experiment results are summarized in Fig.~\ref{fig:scale}, where we have visualized the scalability in terms of the polynomial time complexity with respect to $d$. It can be seen that the computational complexity of our solver is only $\mathcal{O}(d^{0.59})$, which is much smaller than the other schemes. In particular, Gurobi has the highest computational complexity $\mathcal{O}(d^{2.18})$. The computational complexity of the numerical methods and simulated quantum annealing is around $\mathcal{O}(d^{2.09})$, suggesting that they have similar scalability performance. DWave-SD is also significantly faster than the non-von Neumann architectures.

\section{Conclusion}\label{sec:conclusion}

In this paper, we proposed a numerical solver for the DIM. A pruning algorithm has been designed which can effectively downsize the Ising model. An effective domain selection method has been proposed to find the optimal solution in the feasible domains by relaxing the discrete linear programming problem into an optimization of parameters in a continuous space with gradient descent.

The experiments on the randomly generated and fundamental optimization problems demonstrate that our solver is fast, efficient, and scalable. With similar accuracy, our solver can be nearly $10$ times faster than the classical Gurobi. Our scheme also shows a clear advantage in speed over the steepest descent solver of D-Wave and the simulated annealing and simulated quantum annealing algorithms of OpenJij. Therefore, our solver is more suitable for speed-preferred applications, which has the potential to be deployed in the near future to solve large-scale optimization problems, and in the meantime serve as a preferred benchmark for evaluating the performance of quantum computing.

\section*{Acknowledgments}
This work was supported by the National Key R\&D Program of China (No. 2022YFB3304700) and the National Natural Science Foundation of China (No. 62173296).


\bibliographystyle{IEEEtran}
\bibliography{references}

\begin{thebibliography}{10}
\providecommand{\url}[1]{#1}
\csname url@samestyle\endcsname
\providecommand{\newblock}{\relax}
\providecommand{\bibinfo}[2]{#2}
\providecommand{\BIBentrySTDinterwordspacing}{\spaceskip=0pt\relax}
\providecommand{\BIBentryALTinterwordstretchfactor}{4}
\providecommand{\BIBentryALTinterwordspacing}{\spaceskip=\fontdimen2\font plus
\BIBentryALTinterwordstretchfactor\fontdimen3\font minus \fontdimen4\font\relax}
\providecommand{\BIBforeignlanguage}[2]{{%
\expandafter\ifx\csname l@#1\endcsname\relax
\typeout{** WARNING: IEEEtran.bst: No hyphenation pattern has been}%
\typeout{** loaded for the language `#1'. Using the pattern for}%
\typeout{** the default language instead.}%
\else
\language=\csname l@#1\endcsname
\fi
#2}}
\providecommand{\BIBdecl}{\relax}
\BIBdecl

\bibitem{ising1925contribution}
E.~Ising, ``Contribution to the theory of ferromagnetism,'' \emph{Z. Phys}, vol.~31, no.~1, pp. 253--258, 1925.

\bibitem{preskill2018quantum}
\BIBentryALTinterwordspacing
J.~Preskill, ``Quantum computing in the nisq era and beyond,'' \emph{Quantum}, vol.~2, p.~79, 2018. [Online]. Available: \url{https://doi.org/10.22331/q-2018-08-06-79}
\BIBentrySTDinterwordspacing

\bibitem{lucas2014ising}
\BIBentryALTinterwordspacing
A.~Lucas, ``Ising formulations of many np problems,'' \emph{Frontiers in physics}, vol.~2, p.~5, 2014. [Online]. Available: \url{https://doi.org/10.3389/fphy.2014.00005}
\BIBentrySTDinterwordspacing

\bibitem{mcmahon2018solve}
\BIBentryALTinterwordspacing
P.~McMahon, ``To solve optimization problems, just add lasers: An odd device known as an optical ising machine could untangle tricky logistics,'' \emph{IEEE Spectrum}, vol.~55, no.~12, pp. 42--47, 2018. [Online]. Available: \url{https://doi.org/10.1109/mspec.2018.8544983}
\BIBentrySTDinterwordspacing

\bibitem{weinberg2023supply}
\BIBentryALTinterwordspacing
S.~J. Weinberg, F.~Sanches, T.~Ide, K.~Kamiya, and R.~Correll, ``Supply chain logistics with quantum and classical annealing algorithms,'' \emph{Scientific Reports}, vol.~13, no.~1, p. 4770, 2023. [Online]. Available: \url{https://doi.org/10.1038/s41598-023-31765-8}
\BIBentrySTDinterwordspacing

\bibitem{bao2021multi}
\BIBentryALTinterwordspacing
S.~Bao, M.~Tawada, S.~Tanaka, and N.~Togawa, ``Multi-day travel planning using ising machines for real-world applications,'' in \emph{2021 IEEE International Intelligent Transportation Systems Conference (ITSC)}.\hskip 1em plus 0.5em minus 0.4em\relax IEEE, 2021, pp. 3704--3709. [Online]. Available: \url{https://doi.org/10.1109/itsc48978.2021.9564593}
\BIBentrySTDinterwordspacing

\bibitem{pan2023ising}
\BIBentryALTinterwordspacing
Z.~Pan, A.~Sharma, J.~Y.-C. Hu, Z.~Liu, A.~Li, H.~Liu, M.~Huang, and T.~Geng, ``Ising-traffic: Using ising machine learning to predict traffic congestion under uncertainty,'' in \emph{Proceedings of the AAAI Conference on Artificial Intelligence}, vol.~37, no.~8, 2023, pp. 9354--9363. [Online]. Available: \url{https://doi.org/10.1609/aaai.v37i8.26121}
\BIBentrySTDinterwordspacing

\bibitem{hussain2020optimal}
\BIBentryALTinterwordspacing
H.~Hussain, M.~B. Javaid, F.~S. Khan, A.~Dalal, and A.~Khalique, ``Optimal control of traffic signals using quantum annealing,'' \emph{Quantum Information Processing}, vol.~19, pp. 1--18, 2020. [Online]. Available: \url{https://doi.org/10.1007/s11128-020-02815-1}
\BIBentrySTDinterwordspacing

\bibitem{inoue2021traffic}
\BIBentryALTinterwordspacing
D.~Inoue, A.~Okada, T.~Matsumori, K.~Aihara, and H.~Yoshida, ``Traffic signal optimization on a square lattice with quantum annealing,'' \emph{Scientific reports}, vol.~11, no.~1, p. 3303, 2021. [Online]. Available: \url{https://doi.org/10.1038/s41598-021-82740-0}
\BIBentrySTDinterwordspacing

\bibitem{venturelli2015quantum}
\BIBentryALTinterwordspacing
D.~Venturelli, S.~Mandr{\`a}, S.~Knysh, B.~O’Gorman, R.~Biswas, and V.~Smelyanskiy, ``Quantum optimization of fully connected spin glasses,'' \emph{Physical Review X}, vol.~5, no.~3, p. 031040, 2015. [Online]. Available: \url{https://doi.org/10.1103/physrevx.5.031040}
\BIBentrySTDinterwordspacing

\bibitem{parizy2022cardinality}
\BIBentryALTinterwordspacing
M.~Parizy, P.~Sadowski, and N.~Togawa, ``Cardinality constrained portfolio optimization on an ising machine,'' in \emph{2022 IEEE 35th International System-on-Chip Conference (SOCC)}.\hskip 1em plus 0.5em minus 0.4em\relax IEEE, 2022, pp. 1--6. [Online]. Available: \url{https://doi.org/10.1109/socc56010.2022.9908082}
\BIBentrySTDinterwordspacing

\bibitem{venturelli2019reverse}
\BIBentryALTinterwordspacing
D.~Venturelli and A.~Kondratyev, ``Reverse quantum annealing approach to portfolio optimization problems,'' \emph{Quantum Machine Intelligence}, vol.~1, no. 1-2, pp. 17--30, 2019. [Online]. Available: \url{https://doi.org/10.1007/s42484-019-00001-w}
\BIBentrySTDinterwordspacing

\bibitem{benedetti2016estimation}
\BIBentryALTinterwordspacing
M.~Benedetti, J.~Realpe-G{\'o}mez, R.~Biswas, and A.~Perdomo-Ortiz, ``Estimation of effective temperatures in quantum annealers for sampling applications: A case study with possible applications in deep learning,'' \emph{Physical Review A}, vol.~94, no.~2, p. 022308, 2016. [Online]. Available: \url{https://doi.org/10.1103/physreva.94.022308}
\BIBentrySTDinterwordspacing

\bibitem{glover2006handbook}
\BIBentryALTinterwordspacing
F.~W. Glover and G.~A. Kochenberger, \emph{Handbook of metaheuristics}.\hskip 1em plus 0.5em minus 0.4em\relax Springer Science \& Business Media, 2006, vol.~57. [Online]. Available: \url{https://doi.org/10.1007/b101874}
\BIBentrySTDinterwordspacing

\bibitem{bohm2022noise}
\BIBentryALTinterwordspacing
F.~B{\"o}hm, D.~Alonso-Urquijo, G.~Verschaffelt, and G.~Van~der Sande, ``Noise-injected analog ising machines enable ultrafast statistical sampling and machine learning,'' \emph{Nature Communications}, vol.~13, no.~1, p. 5847, 2022. [Online]. Available: \url{https://doi.org/10.1038/s41467-022-33441-3}
\BIBentrySTDinterwordspacing

\bibitem{farhi2001quantum}
\BIBentryALTinterwordspacing
E.~Farhi, J.~Goldstone, S.~Gutmann, J.~Lapan, A.~Lundgren, and D.~Preda, ``A quantum adiabatic evolution algorithm applied to random instances of an np-complete problem,'' \emph{Science}, vol. 292, no. 5516, pp. 472--475, 2001. [Online]. Available: \url{https://doi.org/10.1126/science.1057726}
\BIBentrySTDinterwordspacing

\bibitem{johnson2011quantum}
\BIBentryALTinterwordspacing
M.~W. Johnson, M.~H. Amin, S.~Gildert, T.~Lanting, F.~Hamze, N.~Dickson, R.~Harris, A.~J. Berkley, J.~Johansson, P.~Bunyk \emph{et~al.}, ``Quantum annealing with manufactured spins,'' \emph{Nature}, vol. 473, no. 7346, pp. 194--198, 2011. [Online]. Available: \url{https://doi.org/10.1038/nature10012}
\BIBentrySTDinterwordspacing

\bibitem{albash2018adiabatic}
\BIBentryALTinterwordspacing
T.~Albash and D.~A. Lidar, ``Adiabatic quantum computation,'' \emph{Reviews of Modern Physics}, vol.~90, no.~1, p. 015002, 2018. [Online]. Available: \url{https://doi.org/10.1103/revmodphys.90.015002}
\BIBentrySTDinterwordspacing

\bibitem{wang2013coherent}
\BIBentryALTinterwordspacing
Z.~Wang, A.~Marandi, K.~Wen, R.~L. Byer, and Y.~Yamamoto, ``Coherent ising machine based on degenerate optical parametric oscillators,'' \emph{Physical Review A}, vol.~88, no.~6, p. 063853, 2013. [Online]. Available: \url{https://doi.org/10.1103/physreva.88.063853}
\BIBentrySTDinterwordspacing

\bibitem{inagaki2016coherent}
\BIBentryALTinterwordspacing
T.~Inagaki, Y.~Haribara, K.~Igarashi, T.~Sonobe, S.~Tamate, T.~Honjo, A.~Marandi, P.~L. McMahon, T.~Umeki, K.~Enbutsu \emph{et~al.}, ``A coherent ising machine for 2000-node optimization problems,'' \emph{Science}, vol. 354, no. 6312, pp. 603--606, 2016. [Online]. Available: \url{https://doi.org/10.1126/science.aah4243}
\BIBentrySTDinterwordspacing

\bibitem{dunjko2016quantum}
\BIBentryALTinterwordspacing
V.~Dunjko, J.~M. Taylor, and H.~J. Briegel, ``Quantum-enhanced machine learning,'' \emph{Physical review letters}, vol. 117, no.~13, p. 130501, 2016. [Online]. Available: \url{https://doi.org/10.1103/physrevlett.117.130501}
\BIBentrySTDinterwordspacing

\bibitem{ohzeki2020breaking}
\BIBentryALTinterwordspacing
M.~Ohzeki, ``Breaking limitation of quantum annealer in solving optimization problems under constraints,'' \emph{Scientific reports}, vol.~10, no.~1, p. 3126, 2020. [Online]. Available: \url{https://doi.org/10.1038/s41598-020-60022-5}
\BIBentrySTDinterwordspacing

\bibitem{okada2019improving}
\BIBentryALTinterwordspacing
S.~Okada, M.~Ohzeki, M.~Terabe, and S.~Taguchi, ``Improving solutions by embedding larger subproblems in a d-wave quantum annealer,'' \emph{Scientific reports}, vol.~9, no.~1, p. 2098, 2019. [Online]. Available: \url{https://doi.org/10.1038/s41598-018-38388-4}
\BIBentrySTDinterwordspacing

\bibitem{bohm2018understanding}
\BIBentryALTinterwordspacing
F.~B{\"o}hm, T.~Inagaki, K.~Inaba, T.~Honjo, K.~Enbutsu, T.~Umeki, R.~Kasahara, and H.~Takesue, ``Understanding dynamics of coherent ising machines through simulation of large-scale 2d ising models,'' \emph{Nature communications}, vol.~9, no.~1, p. 5020, 2018. [Online]. Available: \url{https://doi.org/10.1038/s41467-018-07328-1}
\BIBentrySTDinterwordspacing

\bibitem{chowdhury2023full}
\BIBentryALTinterwordspacing
S.~Chowdhury, A.~Grimaldi, N.~A. Aadit, S.~Niazi, M.~Mohseni, S.~Kanai, H.~Ohno, S.~Fukami, L.~Theogarajan, G.~Finocchio \emph{et~al.}, ``A full-stack view of probabilistic computing with p-bits: devices, architectures and algorithms,'' \emph{IEEE Journal on Exploratory Solid-State Computational Devices and Circuits}, 2023. [Online]. Available: \url{https://doi.org/10.1109/jxcdc.2023.3256981}
\BIBentrySTDinterwordspacing

\bibitem{yamaoka201520k}
\BIBentryALTinterwordspacing
M.~Yamaoka, C.~Yoshimura, M.~Hayashi, T.~Okuyama, H.~Aoki, and H.~Mizuno, ``A 20k-spin ising chip to solve combinatorial optimization problems with cmos annealing,'' \emph{IEEE Journal of Solid-State Circuits}, vol.~51, no.~1, pp. 303--309, 2015. [Online]. Available: \url{https://doi.org/10.1109/jssc.2015.2498601}
\BIBentrySTDinterwordspacing

\bibitem{2020A}
\BIBentryALTinterwordspacing
T.~Takemoto, M.~Hayashi, C.~Yoshimura, and M.~Yamaoka, ``A 2 × 30k-spin multi-chip scalable cmos annealing processor based on a processing-in-memory approach for solving large-scale combinatorial optimization problems,'' \emph{IEEE Journal of Solid-State Circuits}, no.~1, p.~55, 2020. [Online]. Available: \url{https://doi.org/10.1109/jssc.2019.2949230}
\BIBentrySTDinterwordspacing

\bibitem{yamamoto2020statica}
\BIBentryALTinterwordspacing
K.~Yamamoto, K.~Kawamura, K.~Ando, N.~Mertig, T.~Takemoto, M.~Yamaoka, H.~Teramoto, A.~Sakai, S.~Takamaeda-Yamazaki, and M.~Motomura, ``Statica: A 512-spin 0.25 m-weight annealing processor with an all-spin-updates-at-once architecture for combinatorial optimization with complete spin--spin interactions,'' \emph{IEEE Journal of Solid-State Circuits}, vol.~56, no.~1, pp. 165--178, 2020. [Online]. Available: \url{https://doi.org/10.1109/isscc19947.2020.9062965}
\BIBentrySTDinterwordspacing

\bibitem{van1987simulated}
\BIBentryALTinterwordspacing
P.~J. Van~Laarhoven and E.~H. Aarts, \emph{Simulated annealing}.\hskip 1em plus 0.5em minus 0.4em\relax Springer, 1987. [Online]. Available: \url{https://doi.org/10.1007/978-94-015-7744-1_2}
\BIBentrySTDinterwordspacing

\bibitem{yamaoka201524}
\BIBentryALTinterwordspacing
M.~Yamaoka, C.~Yoshimura, M.~Hayashi, T.~Okuyama, H.~Aoki, and H.~Mizuno, ``24.3 20k-spin ising chip for combinational optimization problem with cmos annealing,'' in \emph{2015 IEEE International Solid-State Circuits Conference-(ISSCC) Digest of Technical Papers}.\hskip 1em plus 0.5em minus 0.4em\relax IEEE, 2015, pp. 1--3. [Online]. Available: \url{https://doi.org/10.1109/isscc.2015.7063111}
\BIBentrySTDinterwordspacing

\bibitem{aadit2022massively}
\BIBentryALTinterwordspacing
N.~A. Aadit, A.~Grimaldi, M.~Carpentieri, L.~Theogarajan, J.~M. Martinis, G.~Finocchio, and K.~Y. Camsari, ``Massively parallel probabilistic computing with sparse ising machines,'' \emph{Nature Electronics}, vol.~5, no.~7, pp. 460--468, 2022. [Online]. Available: \url{https://doi.org/10.1038/s41928-022-00774-2}
\BIBentrySTDinterwordspacing

\bibitem{wang2024parallel}
\BIBentryALTinterwordspacing
H.~Wang, Z.~Liu, Z.~Xie, L.~Li, Z.~Miao, W.~Cui, and Y.~Pan, ``Parallel ising annealer via gradient-based hamiltonian monte carlo,'' \emph{arXiv preprint arXiv:2407.10205}, 2024. [Online]. Available: \url{https://doi.org/10.48550/arXiv.2407.10205}
\BIBentrySTDinterwordspacing

\bibitem{haribara2016coherent}
\BIBentryALTinterwordspacing
Y.~Haribara, S.~Utsunomiya, and Y.~Yamamoto, ``A coherent ising machine for max-cut problems: performance evaluation against semidefinite programming and simulated annealing,'' \emph{Principles and Methods of Quantum Information Technologies}, pp. 251--262, 2016. [Online]. Available: \url{https://doi.org/10.1007/978-4-431-55756-2_12}
\BIBentrySTDinterwordspacing

\bibitem{rendl2007branch}
\BIBentryALTinterwordspacing
F.~Rendl, G.~Rinaldi, and A.~Wiegele, ``A branch and bound algorithm for max-cut based on combining semidefinite and polyhedral relaxations,'' in \emph{Integer Programming and Combinatorial Optimization: 12th International IPCO Conference, Ithaca, NY, USA, June 25-27, 2007. Proceedings 12}.\hskip 1em plus 0.5em minus 0.4em\relax Springer, 2007, pp. 295--309. [Online]. Available: \url{https://doi.org/10.1007/978-3-540-72792-7_23}
\BIBentrySTDinterwordspacing

\bibitem{veszeli2022mean}
\BIBentryALTinterwordspacing
M.~T. Veszeli and G.~Vattay, ``Mean field approximation for solving qubo problems,'' \emph{Plos one}, vol.~17, no.~8, p. e0273709, 2022. [Online]. Available: \url{https://doi.org/10.1371/journal.pone.0273709}
\BIBentrySTDinterwordspacing

\bibitem{bowles2022quadratic}
\BIBentryALTinterwordspacing
J.~Bowles, A.~Dauphin, P.~Huembeli, J.~Martinez, and A.~Ac{\'\i}n, ``Quadratic unconstrained binary optimization via quantum-inspired annealing,'' \emph{Physical Review Applied}, vol.~18, no.~3, p. 034016, 2022. [Online]. Available: \url{https://doi.org/10.1103/physrevapplied.18.034016}
\BIBentrySTDinterwordspacing

\bibitem{li2023simulated}
\BIBentryALTinterwordspacing
L.~Li, H.~Wang, Z.~Xie, Z.~Liu, W.~Cui, and Y.~Pan, ``Simulated ising annealing algorithm with gaussian augmented {H}amiltonian monte carlo,'' in \emph{2023 42nd Chinese Control Conference (CCC)}.\hskip 1em plus 0.5em minus 0.4em\relax IEEE, 2023, pp. 6760--6765. [Online]. Available: \url{https://doi.org/10.23919/ccc58697.2023.10240679}
\BIBentrySTDinterwordspacing

\bibitem{bulut2021complexity}
\BIBentryALTinterwordspacing
A.~Bulut and T.~K. Ralphs, ``On the complexity of inverse mixed integer linear optimization,'' \emph{SIAM Journal on Optimization}, vol.~31, no.~4, pp. 3014--3043, 2021. [Online]. Available: \url{https://doi.org/10.1137/20m1377369}
\BIBentrySTDinterwordspacing

\bibitem{johnson1995continuous}
\BIBentryALTinterwordspacing
N.~L. Johnson, S.~Kotz, and N.~Balakrishnan, \emph{Continuous univariate distributions, volume 2}.\hskip 1em plus 0.5em minus 0.4em\relax John wiley \& sons, 1995, vol. 289. [Online]. Available: \url{https://doi.org/10.2307/1270425}
\BIBentrySTDinterwordspacing

\bibitem{kingma2014adam}
\BIBentryALTinterwordspacing
D.~P. Kingma and J.~Ba, ``Adam: A method for stochastic optimization,'' \emph{arXiv preprint arXiv:1412.6980}, 2014. [Online]. Available: \url{https://doi.org/10.48550/arXiv.1412.6980}
\BIBentrySTDinterwordspacing

\bibitem{king2015benchmarking}
\BIBentryALTinterwordspacing
J.~King, S.~Yarkoni, M.~M. Nevisi, J.~P. Hilton, and C.~C. McGeoch, ``Benchmarking a quantum annealing processor with the time-to-target metric,'' \emph{arXiv preprint arXiv:1508.05087}, 2015. [Online]. Available: \url{https://doi.org/10.48550/arXiv.1508.05087}
\BIBentrySTDinterwordspacing

\bibitem{hamerly2019experimental}
\BIBentryALTinterwordspacing
R.~Hamerly, T.~Inagaki, P.~L. McMahon, D.~Venturelli, A.~Marandi, T.~Onodera, E.~Ng, C.~Langrock, K.~Inaba, T.~Honjo \emph{et~al.}, ``Experimental investigation of performance differences between coherent ising machines and a quantum annealer,'' \emph{Science advances}, vol.~5, no.~5, p. eaau0823, 2019. [Online]. Available: \url{https://doi.org/10.1126/sciadv.aau0823}
\BIBentrySTDinterwordspacing

\bibitem{oshiyama2022benchmark}
\BIBentryALTinterwordspacing
H.~Oshiyama and M.~Ohzeki, ``Benchmark of quantum-inspired heuristic solvers for quadratic unconstrained binary optimization,'' \emph{Scientific reports}, vol.~12, no.~1, p. 2146, 2022. [Online]. Available: \url{https://doi.org/10.1038/s41598-022-06070-5}
\BIBentrySTDinterwordspacing

\end{thebibliography}


 




\vfill

\end{document}